\documentclass[twocolumn,showpacs,preprintnumbers,amsmath,amssymb,pra]{revtex4-1}
\usepackage{graphicx}
\usepackage{dcolumn}
\usepackage{bm}
\usepackage{color}

\newcommand{\ket}[1]{\left\vert#1\right\rangle}

\begin{document}

\title{Schmidt-like coherent mode decomposition and spatial intensity correlations of thermal light}
\author{I.B.Bobrov} \author{S.S.Straupe} \email{straups@yandex.ru} \author{E.V.Kovlakov, S.P.Kulik}
\affiliation{Faculty of Physics, M.V.Lomonosov Moscow State
University, Russia}

\date{\today}

\begin{abstract}

We experimentally study the properties of coherent mode decomposition for intensity correlation function of quasi-thermal light. We use the technique of spatial mode selection developed for studying transverse entanglement of photon pairs, and show that it can be extended to characterize classical spatial correlations. Our results demonstrate the existence of a unique for a given thermal source basis of coherent modes, correlated in a way much resembling the Schmidt modes of spatially entangled photons.

\end{abstract}


 \maketitle

\section{Introduction}
Entanglement in spatial degrees of freedom of quantum light, such as pairs of photons generated in spontaneous parametric down-conversion (SPDC), is currently an object of active research. These studies are motivated by applications in quantum information science, where infinite-dimensional Hilbert space of spatial states of photons offers attractive capabilities for high-dimensional quantum state engineering. The effective dimensionality of accessible state space is closely related to degree of spatial entanglement for bipartite states, making its experimental quantification an important task. One of the most successful approaches makes use of a Schmidt decomposition. As stated by Law and Eberly in their seminal work \cite{EberlyPRL04}, a unique basis of coherent modes $\{\ket{u_k},\ket{v_k}\}$ exist, such that a spatial state of biphoton pairs $\ket{\Psi_{12}}=\int{dx_1dx_2\Psi(x_1,x_2)a^\dagger(x_1)a^\dagger(x_2)}\ket{\mathrm{vac}} $ can be decomposed as $\ket{\Psi}=\sum_k{\sqrt{\lambda_k}\ket{u_k}\ket{v_k}}$, with these, so-called, Schmidt modes being eigenvectors of reduced single-particle density matrices and $\lambda_k$ - corresponding eigenvalues. A remarkable feature of this decomposition is its single-sum form, implying perfect one-to-one correlations between Schmidt modes. These correlations were studied in several recent works \cite{Straupe_ Ivanov_Kalinkin, MiattoEPJD12, MiattoArXiv12},
and number of significant terms in the decomposition is routinely used as an entanglement quantifier \cite{EberlyPRL04, TorresPRA03, WoerdmanPRA06,
FedorovPRL07, vanExterPRA07, WoerdmanPRL08, BarnettPRA11, PadgettEPJD12, LofflerPRL12}.

A natural question is whether similar approach may be used to study spatial correlations of other origin, for example, correlations in spatially multi-mode classical light? The form of Schmidt decomposition, and the way explicit expressions for Schmidt modes are derived in the case of SPDC biphotons \cite{EberlyPRL04, FedorovJPB09} resembles the well known notion of a coherent mode decomposition for spatial correlation functions in classical statistical optics \cite{WolfJOSA82, WolFJMO93, MandelWolf}. Here we show, that this similarity does not only reside in the form of mathematical equations, but in underlying physics as well. Namely, for a given source of light with quasi-thermal statistics there exist a unique physically distinguished set of coherent modes, correlated in much the same way as Schmidt modes of SPDC. Similar experimental settings would provide similar correlations, with only a quantitative difference in visibility. The situation here resembles that with ghost imaging, which was for some time believed to be possible only due to entanglement \cite{Pittman_Shih} and, thus only with quantum light. It is however well understood now, that spatial correlations required to obtain ghost images should not necessarily be of quantum origin, and a purely classical thermal light may be used as well \cite{LugiatoPRL04, ShapiroPRA08}. In the present work we try to push this analogy further and show, that a framework of Schmidt decomposition may also be generalized to classical states of light.

The paper is organized as follows: in section II we review some properties of coherent mode decomposition for a Gaussian Shell-model of quasi-thermal light and derive the expressions for measured correlation functions, experimental realization is described in section III, and we end with discussion of results and conclusion in section IV.

\section{Coherent mode decomposition for quasi-thermal light}

Spatial coherence properties of classical light are described by coherence function $G^{(1)}(\mathbf{r}_1,\mathbf{r}_2)=\langle E^*(\mathbf{r}_1)E(\mathbf{r}_2) \rangle$, where we neglect the polarization and use scalar complex amplitudes $E(\mathbf{r})$ to describe electromagnetic field. One of the results of classical coherence theory states that it can be decomposed in series of coherent mode functions as
\begin{equation}
\label{WSum}
G^{(1)}(\mathbf{r}_1,\mathbf{r}_2)=\sum_n\lambda_n\phi^{*}_n(\mathbf{r}_1)\phi_n(\mathbf{r}_2),
\end{equation}
where $\phi_n(\mathbf{r})$ are eigenfunctions of an integral operator with a kernel $G^{(1)}(\mathbf{r}_1,\mathbf{r}_2)$ and $\lambda_n$ are corresponding eigenvalues \cite{MandelWolf}.

Higher order correlation functions, such as the intensity correlation function $G^{(2)}(\mathbf{r}_1,\mathbf{r}_2)=\langle I(\mathbf{r}_1)I(\mathbf{r}_2) \rangle$, may be accessed via Hanbury Brown-Twiss (HBT) interferometry, which is exactly the experimental setting commonly used for studying spatial entanglement of photon pairs. In the biphoton case the fourth-order correlation function $G^{(2)}(\mathbf{r}_1,\mathbf{r}_2)$ is measured by counting coincidences between photocounts of two single-photon detectors positioned at points $\mathbf{r}_1$ and $\mathbf{r}_2$, respectively:  $R_c\propto\langle a^\dagger(\mathbf{r}_1)a^\dagger(\mathbf{r}_2)a(\mathbf{r}_1)a(\mathbf{r}_2) \rangle=\left|\Psi(\mathbf{r}_1,\mathbf{r}_2)\right|^2$, with $a(\mathbf{r})$ being the photon annihilation operator and  $\Psi(\mathbf{r}_1,\mathbf{r}_2)$ - the biphoton amplitude. A direct analogue of coherent mode decomposition (\ref{WSum}) for the biphoton amplitude is exactly the Schmidt decomposition. To see whether this analogy goes beyond simply similar mathematical expressions, let us consider the behavior of classical light field in the HBT scheme with spatial mode filters in the arms of the interferometer. Such filters are usually composed of phase holograms followed by single mode fibers and bucket detectors as shown in Fig. \ref{MainSetup}. The main feature of Schmidt decomposition - its single sum form, manifests itself in absence of photocounts in such setup when appropriate mode filters are used to select orthogonal Schmidt modes in different arms of the setup \cite{Straupe_ Ivanov_Kalinkin}. We will show below, that similar effects may be observed with classical quasi-thermal light.

For quasi-thermal light the intensity correlation function $G^{(2)}(\mathbf{r}_1,\mathbf{r}_2)$ can be expressed in terms of the second-order correlation function $G^{(1)}(\mathbf{r}_1,\mathbf{r}_2)$ using Siegert relation:
\begin{equation}
G^{(2)}(\mathbf{r}_1,\mathbf{r}_2)=\langle I(\mathbf{r}_1)\rangle\langle I(\mathbf{r}_2)\rangle+\vert{G^{(1)}(\mathbf{r}_1,\mathbf{r}_2)}\vert^2.
\end{equation}
If the spatial mode filters in the two arms of the HBT scheme are described by impulse response functions $h_1(\mathbf{r}_1,\mathbf{r}'_1)$ and $h_2(\mathbf{r}_2,\mathbf{r}'_2)$, where coordinates $\mathbf{r}$ corresponds to the source plane and $\mathbf{r}'$ - to the detection plane, the transformed intensity correlation function takes the form \cite{LugiatoPRL04, LugiatoPRA2006}:
\begin{eqnarray}
\label{GThermal}
&G^{(2)}(\mathbf{r}'_1,\mathbf{r}'_2) =   \langle \widetilde{I}(\mathbf{r}'_1)\rangle\langle \widetilde{I}(\mathbf{r}'_2)\rangle
 \nonumber\\ &+ \left| \int d \mathbf{r}_1 \int d \mathbf{r}_2 h^{*}_1(\mathbf{r}_1,\mathbf{r}'_1) h_2(\mathbf{r}_2,\mathbf{r}'_2)\right. {}
 \left. G^{(1)}(\mathbf{r}_1,\mathbf{r}_2)  \right|^2,
\end{eqnarray}
where $\widetilde{I}(\mathbf{r}'_{1,2})=\left|\int d\mathbf{r}_{1,2} h_{1,2}(\mathbf{r}_{1,2},\mathbf{r}'_{1,2}) E(\mathbf{r}_{1,2})\right|^2$ is the intensity distribution in the detection plane. One can always choose the mode filters to satisfy the relation:
\begin{eqnarray}
\label{Propagators}
\int d\mathbf{r}_1 h^{(m)}_1(\mathbf{r}_1,\mathbf{r}'_1)\phi_n(\mathbf{r}_1)&\propto& e^{-\frac{(\mathbf{r}'_1)^2}{w_f^2}}\delta_{nm}, \nonumber\\
\int d\mathbf{r}_1 h^{(k)}_2(\mathbf{r}_2,\mathbf{r}'_2)\phi_n(\mathbf{r}_2)&\propto& e^{-\frac{(\mathbf{r}'_2)^2}{w_f^2}}\delta_{nk},
\end{eqnarray}
where $w_f$ is the waist of the fundamental Gaussian mode of the fiber. This choice of propagators corresponds to "projection" on modes $\phi_m(\mathbf{r}_1)$ and $\phi_k(\mathbf{r}_2)$. Substituting (\ref{WSum}) and (\ref{Propagators}) into (\ref{GThermal}) and taking into account that bucket detectors placed after the single mode fiber integrate the signal over the whole detection plane, we obtain the following expression for coincidence to single counts ratio:
\begin{eqnarray}
\label{GThermal_bucket}
\frac{R_c^{(m,k)}}{R_s^{(m)}R_s^{(k)}}\propto g^{(2)}_{(m,k)}\propto 1 + \delta_{mk},
\end{eqnarray}
where
$$g^{(2)}_{(m,k)}=\frac{\int d\mathbf{r}'_1d\mathbf{r}'_2 G^{(2)}(\mathbf{r}'_1,\mathbf{r}'_2)}{\int d\mathbf{r}'_1\langle I(\mathbf{r}'_1)\rangle \int d\mathbf{r}'_2 \langle I(\mathbf{r}'_2)\rangle}$$
is the normalized second order correlation function. It is clear from (\ref{GThermal_bucket}), that spatial correlations between appropriately chosen coherent modes of thermal light show the same correlation features as Schmidt modes of entangled photons - namely, they are only pairwise correlated. The difference turns out to be rather quantitative, than qualitative, similar to the situation with thermal and quantum ghost imaging - the achievable visibility defined as $V=(g^{(2)}_{(m,m)}-g^{(2)}_{(m,n)})/(g^{(2)}_{(m,m)}+g^{(2)}_{(m,n)}),\quad \forall m\neq n$ can not exceed $1/3$ for the thermal light, while it can reach values arbitrarily close to unity for biphotons.

Let us now derive the explicit expressions for coherent modes constituting the decomposition (\ref{WSum}) for a simple model of partially spatially coherent thermal light. Following \cite{WolfJOSA82} we describe light from the quasi-thermal source with a Gaussian Shell-model for the second order coherence function:
\begin{equation}
\label{g1}
G^{(1)}(x_1,x_2)=\sqrt{I(x_1,\omega)I(x_2,\omega)}\mu(x_1-x_2,\omega),
\end{equation}
where spectral intensity distribution and degree of spatial coherence are taken in the form
\begin{equation}
I(x,\omega)=A(\omega)\exp[-x^2/2\sigma_I^2(\omega)],
\end{equation}
\begin{equation}
\label{CoherentF}
\mu(x_1-x_2,\omega)=\exp[-(x_1-x_2)^2/2\sigma_\mu^2(\omega)],
\end{equation}
with $x_1,\ x_2$ being the transverse coordinates, $A(\omega)$ a positive normalization constant, $\sigma_I(\omega)$ - a beam waste and $\sigma_\mu(\omega)$ - its coherence radius. For the sake of simplicity we will first consider a one-dimensional source, generalization to an experimentally relevant two-dimensional case is straightforward. We will also assume the light to be monochromatic and omit the frequency dependence.

In the case of partially spatial coherent light described by a Gaussian Shell-model, one can find the eigenfunctions in a closed analytical form \cite{WolfJOSA82, MandelWolf}:
\begin{equation}
\label{EFumc}
\phi_n(x) = \left(\frac{2c}{\pi} \right)^{1/4}\frac{1}{\sqrt{2^n n!}}H_n(x\sqrt{2c}) \exp(-cx^2),
\end{equation}
\begin{equation}
\label{Lamb1}
\lambda_n=A\left(\frac{\pi}{a+b+c}\right)^{1/2}\left(\frac{b}{a+b+c}\right)^{n},
\end{equation}
where
\begin{equation}
\label{c_parameter}
a=\frac{1}{4 \sigma^{2}_I}, \ b=\frac{1}{4 \sigma^{2}_\mu}, \ c=\sqrt{a^2+2 ab},
\end{equation}
and $H_n(x)$ are Hermite polynomials. Eigenvalues in this coherent mode decomposition are determined by the ratio of coherence radius to beam waist  $\beta=\frac{\sigma_\mu}{\sigma_I}$:
\begin{equation}
\label{lambda_from_beta}
\frac{\lambda_n}{\lambda_0}=\left(\frac{1}{\beta^2 /2+1+\beta[(\beta /2)^2+1]^{1/2}}\right)^n.
\end{equation}
In this simple, but experimentally relevant model the eigenmodes turn out to be familiar Hermite-Gaussian (HG) modes, which were also found to be Schmidt modes of SPDC biphotons under a double-Gaussian approximation for the biphoton amplitude \cite{Straupe_ Ivanov_Kalinkin, FedorovJPB09, MiattoEPJD12}. We note, that similar decomposition may be carried out in polar coordinates giving rise to coherent modes with orbital angular momentum (OAM) \cite{MukundaJOSAA93, MovillaOptLett01, LiOptComm08}. Spectrum of OAM eigenvalues in this decomposition was measured experimentally in \cite{vanExterOptLett10} with interferometric technique.

\section{Experimental realization}
\begin{figure}[h]
\includegraphics[width=80mm]{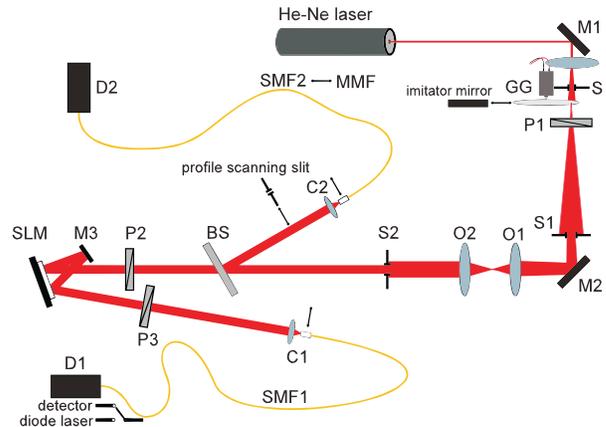}
\parbox{80mm}{\caption{Experimental setup. (See text for detailed description)}\label{MainSetup}}
\end{figure}

The experimental setup is shown in Fig.\ref{MainSetup}. We used a well-known technique to create a pseudothermal light source with controlled spatial coherence \cite{SpillerAJP64}. In this scheme a beam of a He-Ne laser with wavelength $\lambda=632.8 \ \mathrm{nm}$ is scattered on a slowly rotating grounded-glass disc (GG). It is important to make the characteristic time of variation of the scattering pattern larger than the detection time $\tau_{det}\sim2 \mathrm{ns}$, defined by the coincidence circuit time window. The disc was located at the focus of an effective lens consisting of two microscope objectives O1 and O2, forming a collimated beam. Spatial coherence properties of the beam are controlled by iris apertures $S$ and $S_2$, determining the coherence radius $\sigma_\mu$ and the beam waist $\sigma_I$, respectively. The detection part of the setup is an HBT scheme, consisting of a 50/50 non-polarizing beam-splitter; spatial mode filters - a reflective phase-only LCoS SLM in the transmitted channel and a glass step-mask in the reflected one, followed by single-mode fibers $SMF_{1,2}$ placed in the focal planes of 8x microscope objectives $C_{1,2}$; electrical signals from two single-photon detectors $D_{1,2}$ (Perkin-Elmer) were fed to a coincidence curcuit. The model of SLM used was limited to $0.8\pi$ phase shifts only, forcing us to use double-reflection as shown in Fig. \ref{MainSetup}. Polarizers $P_{2,3}$ were used to specify the correct polarization required for operation of polarization sensitive SLM matrix and to remove the undiffracted components, while polarizer $P_1$ was used to control the overall intensity.

The propagator for each arm of the HBT scheme described above is
\begin{widetext}
\begin{equation}
\label{HBTpropagator}
    h(\mathbf{r},\mathbf{r}') \propto \exp\left[-\frac{(\mathbf{r}')^2}{w_f^2} \right] \times \\
    \int d\mathbf{r}{''}\exp(-i \frac{\mathbf{r} \mathbf{r}{''} k}{f}) \times \exp \left[i\mathrm{arg}\left(\mathrm{H}_{m}(\sqrt{2c}x)\mathrm{H}_{n}(\sqrt{2c}y)\right)\right] \times \exp\left[ -\frac{(\mathbf{r}{''}-\mathbf{\mathbf{r}_f})^2}{w^2_f}\right],
\end{equation}
\end{widetext}
where $k=2\pi/ \lambda$ is the wave number, $f$ - the focal length of the microscope objectives $C_{1,2}$, $w_f$ - the waist of the fundamental gaussian mode of the fiber, $\mathbf{r}_f$ - the fiber tip displacement in the focal plane. This propagator defines the detection modes, which for our purposes should be matched with eigenmodes given by (\ref{EFumc}). The mode matching is accomplished by choosing the focal length $ f = \sqrt {\frac {kw^2_f}{4c}}$, in this case the detection mode waist coincides with the waists of the eigenmodes:
$$\frac{k^2 w^2_f} {4f^2} = c=\frac{1}{4\sigma_I^2}\sqrt{1+\frac{2}{\beta^2}}.$$

\begin{figure}[h]
\includegraphics[width=\columnwidth]{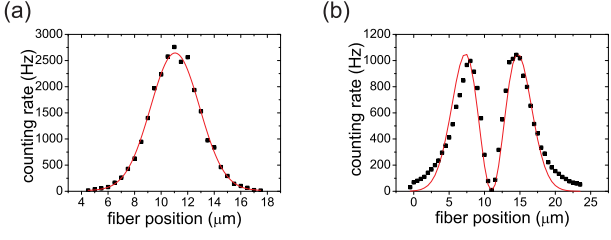}
\parbox{80mm}{\caption{Hermite-Gaussian mode filtering. Convolution of $\mathrm{HG}_{00}$ (a) and $\mathrm{HG}_{10}$ (b) modes with gaussian fiber mode.
Solid red lines are: Gaussian fit used to estimate detection mode waist (a), and $\mathrm{HG}_{10}$ function with the same waist (b).}\label{TEM10conv(imitator)}}
\end{figure}

\begin{figure}[h]
\includegraphics[width=0.8\columnwidth]{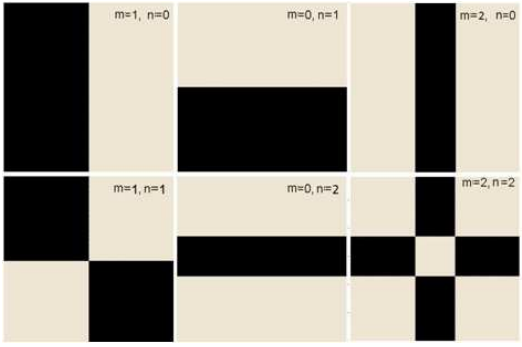}
\parbox{80mm}{\caption{Phase masks corresponding to Hermite-Gaussian modes of different orders. Masks parameters are determined experimentally (see text for details).}\label{HGholo}}
\end{figure}

The mode filters were tested with a coherent single mode laser beam. For this purpose the grounded-glass disc was replaced by a mirror and radiation of a 650 nm diode laser was coupled into $SMF_1$ fiber in "backward" direction (see Fig.\ref{MainSetup}). The after passing through both arms of the setup the beam was detected in the reflected arm. One may consider this procedure to be the direct implementation of Klyshko's "advanced-wave picture" \cite{KlyshkoPLA_1988}. The intensity at the output of the fiber is proportional to the convolution of $\mathrm{HG}_{mn}$ and $\mathrm{HG}_{00}$ modes
\begin{equation}
I^m_2(x_{f}) \sim \left| \int^{\infty}_{-\infty} \mathrm{HG}_{m0}(\xi)\mathrm{HG}_{00}(x_f-\xi) d\xi \right|^2,
\end{equation}
where $x_f$ was the horizontal coordinate of the tip of $\mathrm{SMF}_2$. Convolutions for the first two modes $\mathrm{HG}_{00}$ and $\mathrm{HG}_{10}$ are shown in Fig. \ref{TEM10conv(imitator)}. Data shown in the Figure \ref{TEM10conv(imitator)}(a) were used to determine the width of the fiber mode $(w_f=3.69\pm0.04 \ \mathrm{\mu m})$. Phase masks used for mode conversion had step-like phase profiles shown in Fig. \ref{HGholo}. Sizes and positions of the phase masks relative to the beam were determined experimentally following the operational criterion of minimizing the counting rate in the central position of the $\mathrm{SMF}_2$ fiber.

\begin{figure}[h]
\includegraphics[width=\columnwidth]{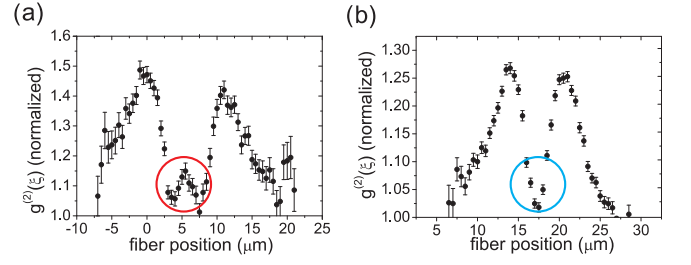}
\parbox{80mm}{\caption{Convolutions of the gaussian detection mode with $HG_{20}$ modes of different waists. (a) The waist $c$ is "incorrect", the mode-matching condition is not satisfied, normalized $g^{(2)}_{(2,0)}(x_f=0)>1$. (b) The waist is chosen appropriately, normalized intensity correlation function reaches the minimal value of unity.}\label{Conv_wrong}}
\end{figure}

Satisfying the condition $\frac{k^2 w^2_f} {4f^2} = c$ is essential to obtain the desired matching of the detected modes and the eigenmodes of coherence function. In the actual experiment the detection mode waist was fixed, while the mode-matching was done by changing the parameters of the input beam. Width of the apertures $\mathrm{S}$ and $\mathrm{S}_2$ determine the coherence radius $\sigma_\mu$ and the beam waist $\sigma_I$, correspondingly, and thus define the eigenmodes width $c$ through (\ref{c_parameter}). One can show by direct calculation that the value of normalized intensity coherence function $g^{(2)}_{(m,0)}(\mathbf{r}_f=0)=1$ for even $m$ only if the mode-matching condition is satisfied, otherwise it is always larger than 1, as shown in Fig. \ref{Conv_wrong}. This fact was used as an operational criterion of good mode-matching. The corresponding beam waist and coherence radius were measured to be $\sigma_I= (2.3\pm0.1)$~mm and $\sigma_\mu=(0.57\pm0.02)$~mm, corresponding to the value of $\beta=0.24\pm0.02$.

\begin{figure}[h]
\includegraphics[width=\columnwidth]{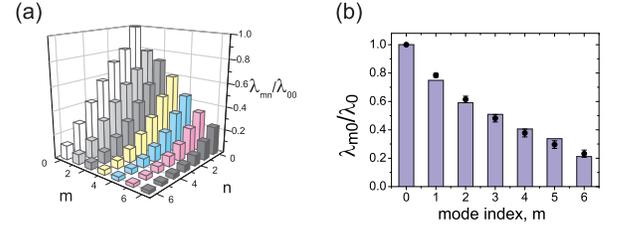}
\parbox{80mm}{\caption{Experimentally measured eigenvalues of coherent mode decomposition. (a) Two-dimensional distribution of normalized eigenvalues $\lambda_{mn}/\lambda_0$ (indexes $m$ and $n$ correspond to $\mathrm{HG}_{m0}$ and $\mathrm{HG}_{0n}$ modes, respectively). (b) One-dimensional section ($n=0$) of the distribution. Blue bars are experimental data, black circles are values calculated from (\ref{lambda_from_beta}) using independently measured value $\beta=0.24\pm0.02$. Error bars correspond to error in measured value of $\beta$, error bars for experimental points are smaller than point markers.}\label{Eigenvalues}}
\end{figure}

With the mode-matching conditions satisfied, the spatial mode filters in both arms of the setup act as projectors on eigenmodes of coherent mode decomposition~(\ref{WSum}). In this case the single counts rate of a detector in one of the arms is proportional to partial intensity of a constituent mode selected by the filter:
\begin{equation}
\label{PartialIntensity}
I_{(m)}\propto \int d\mathbf{r'}\left|\int d\mathbf{r} h_{1,2}^{(m)}(\mathbf{r,r'},)E(\mathbf{r})\right|^2\propto\lambda_m.
\end{equation}
It provides a way to measure the eigenvalues in~(\ref{WSum}) experimentally. The distribution of eigenvalues, normalized to $\lambda_0$ is shown in Fig.~\ref{Eigenvalues}. We find excellent agreement with theoretical predictions obtained from~(\ref{lambda_from_beta}) using the independently measured value $\beta=0.24\pm0.02$.

\begin{figure}[h]
\includegraphics[width=\columnwidth]{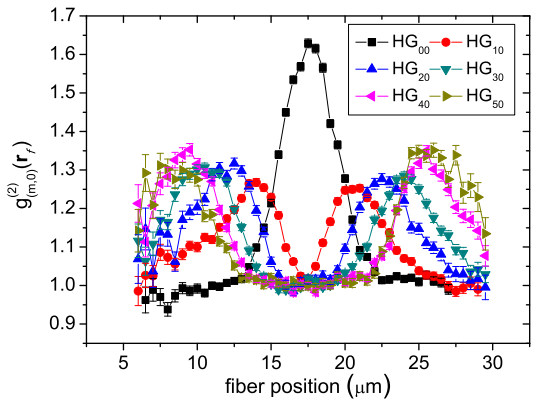}
\parbox{80mm}{\caption{Spatial structure of eigenmodes: dependence of normalized $g^{(2)}_{(m,0)}$ function on fiber displacement in the reflected channel. Unity values for zero displacement correspond to absence of correlations for modes of different order, predicted by theory.}\label{TEMm0Conv}}
\end{figure}

The spatial shape of modes (\ref{EFumc}) selected by corresponding mask in the transmitted channel is revealed in the dependence of intensity correlation function on fiber displacement in the reflected channel. Indeed, if a phase mask corresponding to \emph{m}-th eigenmode is installed in the transmitted channel with no mask in the reflected, using (\ref{GThermal}) one can find the following expression for a normalized intensity correlation function:
\begin{equation}
\label{Convolution}
    g^{(2)}_{(m,0)}(\mathbf{r}_f)=1+\frac{\left|\lambda_m\alpha_m(\mathbf{r}_f)\right|^2}{\lambda_m\sum\limits_k{\left|\alpha_k(\mathbf{r}_f)\sqrt{\lambda_k}\right|^2}},
\end{equation}
where $\mathbf{r}_f$ is the position of the single mode fiber in the reflected channel, and
\begin{equation}
\label{alpha_conv}
    \alpha_k(\mathbf{r}_f)=\int d\mathbf{r} \phi_k(\mathbf{r})\phi_0(\mathbf{r}-\mathbf{r}_f).
\end{equation}
Experimental data for $\mathrm{HG}_{m0}$ modes are shown in Fig.~\ref{TEMm0Conv}. It is clear, that the selected modes are indeed orthogonal in the sense, that $g^{(2)}_{(m,0)}(0)=1,\quad \forall m\neq0$, consistently with (\ref{GThermal_bucket}). Dependence on the fiber displacement is also in qualitative agreement with (\ref{Convolution}). However for large displacements we observe anomalously high values of $g^{(2)}_{(m,0)}(\mathbf{r}_f)$ for large $m$. This is most probably an artefact of our method for mode selection based on using phase-only step masks, inevitably causing higher-order mode contributions and compromising mode purity for large displacements.

\begin{figure}[h]
\includegraphics[width=\columnwidth]{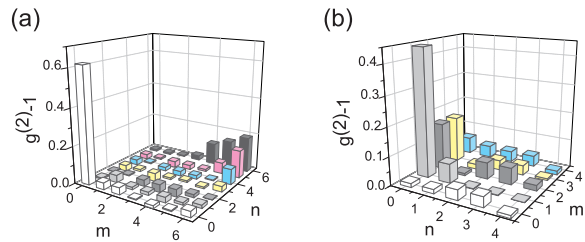}
{\caption{Mode cross-correlations: (a) $g^{(2)}_{(m,0)}-1$ values measured with $\mathrm{HG}_{m0}$ modes selected in the transmitted arm and $\mathrm{HG}_{00}$ mode in the reflected one; (b) $g^{(2)}_{(m,1)}-1$ values, corresponding to $\mathrm{HG}_{m0}$ modes filtered in the transmitted arm, and $\mathrm{HG}_{01}$ mode selected in the reflected arm.}\label{G2mn}}
\end{figure}

For zero displacement and low-order modes we find the detected modes to be close to Hermite-Gaussian eigenmodes, and thus they are pairwise correlated, as illustrated by Fig. \ref{G2mn}. We expect $g^{(2)}_{(mn)}(\mathbf{r}_f=0)$ to behave in accordance with expression (\ref{GThermal_bucket}), i.e. to be essentially unity for modes of different order, and larger than unity only for modes of the same order. Experimental results of Fig. \ref{G2mn} confirm this statement, although some amount of anomalous correlations is observed for modes of high order. These correlations are more pronounced on Fig. \ref{G2mn} (b), which may be partially explained by much poorer quality of the glass phase-masks used in the reflected channel. Nevertheless, one can clearly observe the good agreement with expected correlation properties of quasi-Schmidt coherent modes.

\section{Discussion and conclusion}

\begin{figure}[h]
\includegraphics[width=\columnwidth]{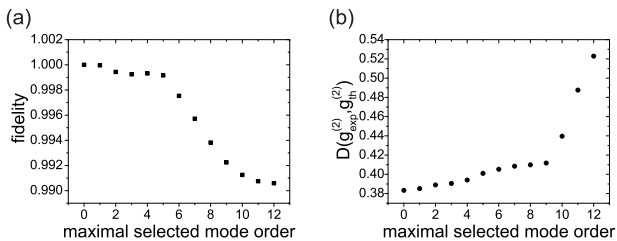}
{\caption{(a) Quality of experimentally obtained distribution of eigenvalues. Dependence of fidelity calculated according to expression (\ref{fidelity}) on the maximal order of modes taken into account $M+N$. (b) Similar dependence of vector-norm distance between experimental and theoretical values of $g^{(2)}_{(m,n)}$, characterizing cross-correlations in higher-order modes.}\label{FidelityEval}}
\end{figure}

The main source of discrepancies between our experimental results and theoretical expectations for ideal projective measurements in Schmidt-like basis is the usage of non-ideal mode filters. Technical limitations of the SLM used do not allow to use it for amplitude modulation, which is necessary for perfect transformations of Hermite-Gaussian modes. Nevertheless, our results are very similar to what is expected for exact Hermite-Gaussian filters. Let us address this issue quantitatively. One can define \emph{fidelity}
\begin{equation}
\label{fidelity}
F=\frac{\sum\limits_{m,n}^{N,M}{ \sqrt{\lambda^{th}_{mn}\lambda^{exp}_{mn}}}} {\sqrt{\left(\sum\limits_{m,n}^{M,N}{\lambda^{th}_{mn} } \right)}\sqrt{\left(\sum\limits_{m,n}^{M,N}{\lambda^{exp}_{mn}} \right)}},
\end{equation}
as a measure of deviations of (normalized) distribution of experimentally obtained eigenvalues $\lambda_{mn}^{exp}$ from the product of two (normalized) theoretical distributions (\ref{lambda_from_beta}) for a two-dimensional Gaussian Shell-model of incoherent light. The value of this quantity for our data is dependent on the upper limits $\left\{M,N \right\}$, i.e. on the maximal order of modes, that are taken into account. This dependence is shown in Fig.~\ref{FidelityEval} (a). As expected the fidelity decreases when values with larger indexes are taken into account, since the step-like phase holograms performance is worse for higher-order modes. Nevertheless, values of $F>0.99$ indicate excellent agreement with theoretical distribution. To estimate the amount of excess correlations between modes with different indexes, leading to $g^{(2)}_{(m,n)}>1$ in Fig. \ref{G2mn}, one can use any appropriate distance measure. We choose to use vector-norm distance $D(g^{(2)}_{exp},g^{(2)}_{th})= \sqrt{\sum_{m,n}{ \left(g^{(2)}_{exp, (m,n)}-g^{(2)}_{th, (m,n)}\right)^2}}$. Dependence plotted in Fig. \ref{FidelityEval} (b) also shows deviation from expected values for higher-order modes.

In conclusion, we have investigated spatial intensity correlations in partially spatially coherent thermal light using the technique of projective measurements originally developed to study spatial entanglement of photon pairs. In particular, our detection scheme is similar to those used in experiments on orbital angular momentum entanglement \cite{TornerNaturePhys07} and essentially the same as was used by authors to study approximate Schmidt decomposition for SPDC biphotons \cite{Straupe_ Ivanov_Kalinkin}. We have found much similarity in properties of coherent mode decomposition for intensity correlation function of classical thermal light and Schmidt decomposition for biphoton amplitude. Although, the expressions for eigenmodes and eigenvalues are different, they are still governed by a single experimental parameter $\beta=\sigma_\mu/\sigma_I$ - the ratio of coherence radius to the width of intensity distribution. A similar quantity for entangled photons is known as the Fedorov ratio $R=\Delta x_c/\Delta x_s$ - the ratio of widths of coincidence (conditional) and single counts (single particle) distributions in either coordinate or momentum space of photons \cite{FedorovPRA05}. This quantity is shown to be equal to Schmidt number $K$, which is the effective number of non-zero eigenvalues $\lambda_i$ in the Schmidt decomposition, defined as $K=\sum_i{\lambda_i^{-2}}$. Thus for a pure two-photon state it can serve as a measure of spatial entanglement. It is clear, that this quantity has a purely classical analogue - optical etendue \cite{WoerdmanPRA06}. Results of this work show, that analogy between spatial entanglement and classical spatial correlations goes further. In purely operational terms the qualitative similarity is almost complete, with the difference being only quantitative - excess of $g^{(2)}$ over unity is limited to 1, while it can be unlimited in the case of entangled photons.

One can consider the described experiments as a special case of thermal light ghost interference/imaging with mode-sensitive detection. In this context our work adds new arguments to the discussion of quantum and classical features of spatial correlations underlying this method. The presented results may contribute to deeper understanding of what is classical and what is really quantum in spatial correlations of multi-mode light.

We are grateful to A.~A.~Kalinkin for considerable help at the early stage of the experiment. This work has been founded by RFBR grants 10-02-00204, 12-02-00288 and 12-02-31041, and the Federal Program of the Russian Ministry of Education and Science (grant 8393). S.~S.~Straupe is grateful to the "Dynasty" foundation for financial support.

\end{document}